\documentstyle[12pt]{article}
\newcommand{\doublespace}{
   \renewcommand{\baselinestretch}{1.5}
   \large\normalsize}

\renewcommand{\u}[1]{{\bf #1}}
\newcommand{\be}{\begin{equation}}
\newcommand{\by}{\begin{eqnarray}}
\newcommand{\ee}{\end{equation}}
\newcommand{\ey}{\end{eqnarray}}

\newcommand{\Sp}{\makebox[.5in]{}}
\newcommand{\ra}{\rightarrow}

\renewcommand{\thesubsection}{(\Alph{subsection})}
\renewcommand{\mathbf}{\bf}
\begin{document}
\doublespace

\begin{center}
\large {\bf  Optimized RVB states of the 2-d antiferromagnet:\\
Ground state and excitation spectrum \\}
\normalsize  Yong-Cong  Chen and Kai Xiu\\
Department of Physics\\
University of Science and Technology of China\\
Hefei, Anhui 230026, China \\
 Fax: 011-86-551-331760; Tel: 011-86-551-301075\\
\vspace{.15in}
\end{center}

\newcommand{\bk}{{\mathbf k}}
\begin{center} {\bf Abstract}\end{center}
The Gutzwiller projection of the Schwinger-boson mean-field
solution of the 2-d spin-1/2 antiferromagnet in a square lattice is shown to
produce the optimized, parameter-free RVB ground state. We get $-0.6688J$/site
and $0.311$ for the energy and the staggered magnetization. The
spectrum of the excited states is found to be linear and gapless near $\bk\cong
0$.  Our calculation suggests, upon breaking of the rotational symmetry,
$\epsilon_{\bk}\cong 2JZ_{r} \sqrt{1-\gamma_{\bk}^{2}}$ with
$Z_{r}\cong 1.23$.
\vspace{.1in}

\vspace{.1in}

\noindent KEY words: RVB state, Schwinger boson,
Heisenberg model, Gutzwiller projection.
\vspace{.1in}

\noindent Running title:  Optimized RVB states of antiferromagnet
\vspace{.5in}

\noindent To appear in Phys. Lett. A (1993).\\
  Current address of Y.-C. Chen: \\
Department of Mathematics, Rutgers University, New Brunswick, NJ
08903; Tel: 908-932-3726; Fax: 908-932-5530; email: ychen@math.rutgers.edu.

\newpage
\renewcommand{\d}{\dagger}
\newcommand{\ua}{\uparrow}
\newcommand{\da}{\downarrow}

It is widely realized that the 2-d antiferromagnetism may play a
central role in the copper-oxide superconductors[\ref{r1}].  A simple yet
useful model for this  problem is the Heisenberg model in a square lattice
with  nearest-neighbor couplings. Although there is no
exact solution yet for the model, the traditional spin-wave theory,
which works at zero temperature with broken rotational symmetry,
has been known to describe quite well its properties[\ref{r3}].
However, away from the half-filling the  long-range correlations are quickly
destroyed. In order to understand the entire behavior of high-T$_{c}$
materials, a truly two-dimensional theory with the potential
of generalizing to the doped regime is highly desired.
The Schwinger-boson approach stands as a good candidate for this
purpose.  It has been shown to produce the qualitatively  or even
quantitatively correct low-temperature behavior of
the Heisenberg model[\ref{r5},\ref{r5a}]. Yet
extension to the general $t-J$ model is rather
straightforward.  For example, several groups have predicted using
this formalism the possible existence of the spiral states at
finite dopings[\ref{r5b}].
 Despite of these apparent successes,  it must be stressed that they are
all mean-field results.  The particle number constraint (having one boson
per site) was treated {\em on average} only.  The constraint which
greatly limits the physical states of the system still presents the main
technical obstacle: the including of many
unphysical states in most of the mean-field approaches
may invalidate the results.

To go beyond the mean-field consideration, we have
proposed in a series of recent work[\ref{r6}-\ref{r8}]
a general scheme for carrying out a complete Gutzwiller
projection on Schwinger-boson mean-field states of the Heisenberg and the
$t-J$ models. {\em The new approach attempts to treat the constraint exactly}.
It has been used to study the corrections, due to the particle number
constraint,  on the mean-field results, and to find  suitable variational
states for the $t-J$ model. But only the static properties were calculated.
In this letter, we shall study the ground state projected from the
Schwinger-boson mean-field solution of the Heisenberg model.
In particular, we shall show the
followings: 1) It is the optimized RVB state of the 2-d antiferromagnet
within an analytic  self-consistency approximation. 2) The
approximation gives $-0.3352J$/bond for the ground-state energy
and $0.303$ for the staggered magnetization, in exact
agreement with the modified  spin-wave theory (to the order $1/S$).
Furthermore, an exact Monte Carlo evaluation yields $-0.3344J$/bond
and $0.311$, much closer to the best estimated values[\ref{r3}]
$-0.3346(1)J$/bond and $0.31(2)$.
This confirms the optimization. 3) The excited states are
explicitly constructed and analyzed. The excitation spectrum is found to be
linear and gapless near $\bk\cong 0$, in agreement with some well-established
results.  The Monte Carlo calculation also yields, upon breaking of the
rotational symmetry,  $\epsilon_{\bk}\cong 2JZ_{r}
\sqrt{1-\gamma_{\bk}^{2}}$ with $Z_{r}\cong 1.23$.
The present study brings useful new ingredients to the RVB states
in the context of high-$Tc$ theories.

In terms of the Schwinger-boson representation,
the nearest-neighbor antiferromagnetic  Heisenberg
model may be expressed as
\be\label{e1}
\hat{H}=J\sum_{<ij>}[-\frac{1}{2}\hat{A}_{ij}^{\d}\hat{A}_{ij}+S^{2}]
\ee
where
\be\label{e2}
\hat{A}_{ij}=\sum_{\sigma=\ua\da}\hat{b}_{i\sigma}\hat{b}_{j\sigma},
\Sp
\sum_{\sigma=\ua\da}\hat{b}^{\d}_{i\sigma}\hat{b}_{i\sigma}=2S.
\ee
In the above, a unitary transformation
$\hat{b}_{j\ua}\ra-\hat{b}_{j\da}$,
$\hat{b}_{j\da}\ra \hat{b}_{j\ua}$ has been performed
on one of the sublattices (say $B$).
The partition function may be calculated via
\[Z(\beta)=\mbox{Tr}[\hat{{\cal P}}_{G}\hat{\rho}]
=\mbox{Tr}\left\{[\prod_{i=1}^{N}\hat{P}_{i}]\hat{\rho}\right\},
\Sp
\hat{\rho}=\exp(-\beta\hat{H})\]
where  $\hat{{\cal P}}_{G}$ ($\hat{P}_{i}$) stands for the
Gutzwiller projection operator for the whole lattice (the $i$th site) which
enforces the constraint in (\ref{e2}).  In [\ref{r7},\ref{r8}], we have
shown that $\hat{P}_{i}$ can be replaced by a differential operator $P_{i}$,
\be\label{e3}
\left.Z(\beta)=\left[\prod_{i=1}^{N}P_{i}\right]<\{\alpha_{i}^{\d}\}|\hat{\rho}
|\{\alpha_{i}\}>\right|_{\{\alpha_{i}^{\d},\alpha_{i}=0\}}
\Sp
P_{i}=\frac{1}{(2S)!}\left[\sum_{\sigma=\ua\da}\frac{\partial}{\partial
b_{i\sigma}}\frac{\partial}{\partial b_{i\sigma}^{\d}}\right]^{2S},
\ee
provided that the matrix elements of $\hat{\rho}$ are known:
 Here $\alpha_{i}=(b_{i\ua},b_{i\da})$ are complex numbers
and $|\alpha_{i}>\equiv
\exp(\sum_{\sigma=\ua\da}\hat{b}_{i\sigma}^{\d}
b_{i\sigma})|0>$ is a usual coherent state.
In the Schwinger-boson mean-field theory, one replaces the
Hamiltonian (\ref{e1}) by a mean-field one,
\be\label{e4}
\hat{H}_{\mbox{mf}}=E_{0}-J\sum_{<ij>,\sigma}[D_{ij}
\hat{b}^{\d}_{i\sigma}\hat{b}^{\d}_{j\sigma}+h.c.]
+\lambda\sum_{i,\sigma}\hat{b}^{\d}_{i\sigma}\hat{b}_{i\sigma}.
\ee
Thus (\ref{e3}) can be used to calculate the effects of the
constraint. To be  definite, we shall restrict ourselves to the 2-d model with
$S=1/2$.

\subsection{The projected ground state}

We now try to project out a ground-state wave function out of the
mean-field solution of (\ref{e4}). This can be done by putting
$\beta\ra\infty$.  The calculation of
$<\{\alpha_{i}^{\d}\}|\hat{\rho}_{\mbox{mf}}|\{\alpha_{i}\}>$
is straightforward.  We obtain[\ref{r8}],
\newcommand{\br}{{\mathbf r}}
\be\label{e5}
<\{\alpha_{i}^{\d}\}|\hat{\rho}_{\mbox{mf}}|\{\alpha_{i}\}>
=Z_{0}(\beta)\exp\left(\sum_{\bk,\sigma}[
W^{(1)}_{\bk}b^{\d}_{\bk\sigma}b_{\bk\sigma}+\frac{1}{2}(W^{(2)}_{\bk}
b^{\d}_{-\bk\sigma}b^{\d}_{\bk\sigma}+c.c.)]\right),
\ee
where
\be\label{e6}
W^{(1)}_{\bk}=
[(\lambda/\omega_{\bk})
\sinh(\beta\omega_{\bk})+\cosh(\beta\omega_{\bk})]^{-1},
\Sp
W^{(2)}_{\bk}=(JD_{\bk}/\omega_{\bk})\sinh(\beta\omega_{\bk})W^{(1)}_{\bk},
\ee
with $\omega_{\bk}=\sqrt{\lambda^{2}-J^{2}|D_{\bk}|^{2}}$.
Here, $b^{\d}_{\bk\sigma}$, $b^{\d}_{\bk\sigma}$  and $D_{\bk}$
are, respectively, the Fourier transforms of $b^{\d}_{i\sigma}$,
$b_{i\sigma}$ and $D_{ij}$. Taking the uniform-bond solution
$D_{ij}=D$ gives $D_{\bk}=zD\gamma_{\bk}$ with $\gamma_{\bk}=[\cos
k_{x}+\cos k_{y}]/2$.  At $\beta=\infty$, $W^{(1)}_{\bk}$ vanishes
and  $\{b^{\d}_{i\sigma}\}$ and $\{b_{i\sigma}\}$
in $<\{\alpha_{i}^{\d}\}|\hat{\rho}_{\mbox{mf}}|\{\alpha_{i}\}>$
are fully decoupled. The ground state is simply
the $\{b^{\d}_{i\sigma}\}$ part of the Gutzwiller projection.
This yields the wave function in the coherent-state
representation,
\be\label{e7}
<\{\alpha^{\d}_{i}\}|\Phi_{G}>=Y_{N}^{-1/2}\times
\left.\left[\prod_{i=1}^{N}(\sum_{\sigma=\ua\da}
b^{\d}_{i\sigma}\frac{\partial}{\partial
\tilde{b}^{\d}_{i\sigma}})\right]
\exp\left(\frac{1}{2}\sum_{i,j,\sigma}W_{ij}\tilde{b}^{\d}_{i\sigma}
\tilde{b}^{\d}_{j\sigma}\right)\right|_{\{\tilde{\alpha}^{\d}_{i}=0\}},
\ee
where $Y_{N}$ is the normalization constant and
[we drop, hereafter, the superscript ``(2)'']
\[W_{ij}=\frac{1}{N}\sum_{\bk} \frac{\eta \gamma_{\bk}
\exp[i\bk\cdot(\br_{i}-\br_{j})]
}{1+\sqrt{1-(\eta\gamma_{\bk})^{2}}}\]
\be\label{e8}
=\frac{1}{N}\sum_{\bk}W_{\bk}\exp[i\bk\cdot(\br_{i}-\br_{j})],
\Sp \eta=\frac{zD}{\lambda}\ra 1.
\ee
Eq. (\ref{e7}) can be easily checked by inserting $\hat{\rho}=
|\Phi_{G}><\Phi_{G}|$ into (\ref{e3}).  The latter
will be employed in the following calculations.
Note that the ground state so obtained is a
{\em truly RVB type}[\ref{r9}], the bond strength is simply
$W_{ij}$. But we shall see that the $|\Phi_{G}>$ possesses
long-range  antiferromagnetic correlations, in agreement with other
theories.

\subsection{The self-consistency approximation}

The evaluation of $Y_{N}$ can be reduced to a generalized
loop gas problem[\ref{r9},\ref{r10}]. A general
procedure was outlined in [\ref{r7},\ref{r8}]. But the
self-consistency approximation needs to be further elucidated.  The effect of
$P_{i}$ can be most easily handled via a $4\times 4$ transfer matrix ${\mathbf
T}_{ij}$ ($i,j$ are the lattice sites). Suppose that, at a given stage, the
relevant part of the prefactor contains $g^{\d}_{i\ua}b_{i\ua}+
g_{i\ua}b_{i\ua}^{\d}+g_{i\da}^{\d}
b_{i\da}+g_{i\da}b_{i\da}^{\d}$ (They are brought down by
previous differentiations.  Prefactors not in this form will start
new loops). Taking $P_{i}$ leads to the prefactor for site $j$ via
\be\label{e9}
\left(\begin{array}{c} {\mathbf g}_{j\ua} \\ {\mathbf g}_{j\da}
\end{array}\right)
=\left(\begin{array}{cc} {\mathbf G}_{ij} & {\mathbf 0} \\ {\mathbf 0} &
{\mathbf G}_{ij}\end{array}\right)
\left(\begin{array}{c} {\mathbf g}_{i\ua} \\ {\mathbf g}_{i\da}
\end{array}\right);
\Sp {\mathbf g}_{i\sigma}=
\left(\begin{array}{c} g_{i\sigma} \\ g^{\d}_{i\sigma}
\end{array}\right),
{\mathbf G}_{ij}=\left(\begin{array}{cc} 0 & W_{ij} \\ W_{ij}^{\ast} &
0\end{array}\right)
\ee
which defines ${\mathbf T}_{ij}$.  We shall refer to the four-component
vectors in (\ref{e9}) as the {\em states}  of the sites.  Tracing over the
matrix of a close loop yields its contribution.  $Y_{N}$ is  then the sum of
all possible configurations of {\em self-avoiding} loops.  To
illustrate this, we can decompose
$Y_{N}$ into (all $j_{k}$'s below are different)
\be\label{e10}
Y_{N}=\sum_{n=0}^{\infty}\left\{\sum_{\{j_{k}; k\neq
0\}}Y_{N-n-1}(\{j_{k}\})
\times \mbox{Tr}[{\mathbf G}_{j_{0}j_{1}}\cdots{\mathbf G}_{j_{n}j_{0}}]
\right\}.
\ee
The arguments of $Y_{N-n-1}$ are the sites excluded.
The spin-spin correlation between {\em different sublattices}
can be similarly calculated.  Substituting
$\hat{b}_{i\sigma}^{\d}\ra
b_{i\sigma}^{\d}$, $\hat{b}_{i\sigma}\ra \partial /\partial
b^{\d}_{i\sigma}$,
we have
\[\hat{{\mathbf S}}_{i}\cdot\hat{{\mathbf S}}_{j}
\longrightarrow -\frac{1}{2}
\sum_{\sigma=\ua\da}b^{\d}_{i\sigma}b^{\d}_{j\sigma}
\frac{\partial }{\partial b^{\d}_{i\sigma}}
\frac{\partial }{\partial b^{\d}_{j\sigma}}
+\frac{1}{4}
-\frac{1}{2}
\sum_{\sigma=\ua\da}b^{\d}_{i\sigma}b^{\d}_{j\sigma}
\frac{\partial }{\partial b^{\d}_{i-\sigma}}
\frac{\partial }{\partial b^{\d}_{j-\sigma}}.\]
One needs to modify the transfer matrices at sites $i$ and $j$
(they are correlated). This can be accomplished by multiplying
simultaneously the following $4\times 4$ matrices
to the {\em states} of $i$ and $j$
\be\label{e11}
-\frac{1}{4}\left(\begin{array}{cc} {\mathbf I} & {\mathbf 0}\\ {\mathbf 0}&
-{\mathbf I}
\end{array}\right)_{i}
\left(\begin{array}{cc} {\mathbf I} & {\mathbf 0}\\ {\mathbf 0}& -{\mathbf I}
\end{array}\right)_{j}
-\frac{1}{2}\left[
\left(\begin{array}{cc} {\mathbf 0} & {\mathbf 0}\\ {\mathbf I}& {\mathbf 0}
\end{array}\right)_{i}
\left(\begin{array}{cc} {\mathbf 0} & {\mathbf I}\\ {\mathbf 0}& {\mathbf 0}
\end{array}\right)_{j}+
\left(\begin{array}{cc} {\mathbf 0} & {\mathbf I}\\ {\mathbf 0}& {\mathbf 0}
\end{array}\right)_{i}
\left(\begin{array}{cc} {\mathbf 0} & {\mathbf 0}\\ {\mathbf I}& {\mathbf 0}
\end{array}\right)_{j}
\right]
\ee
One then can compute the modified $\tilde{Y}_{N}$.
It turns out that only
configurations with $i$ and $j$ on the same loops contribute, which
are $-3/4 $ of those in $Y_{N}$, agreeing with the
rules proposed in [\ref{r9},\ref{r10}].

We now present an analytic self-consistency approximation for
$Y_{N}$ and $<\hat{{\mathbf S}}_{i}\cdot\hat{{\mathbf S}}_{j}>$.  Let us first
approximate $Y_{N}/Y_{N-n-1}\ra y^{n+1}$.
This assigns a uniform weight $1/y^{n+1}$ for a loop of $(n+1)$
bonds when both sides of (\ref{e10}) are divided by $Y_{N}$.
The most difficult part of the problem is the self-avoiding
restriction. Since the system is expected to have long-range correlations,
it may be reasonable to ignore this
restriction at the first place. Denote the resulting $y$ by
$y_{0}$. We then recover the correct $y$ by taking into account the
over-counting.  This allows a full analytic summation over the loops in
(\ref{e10}), which leads to a self-consistency equation for $y_{0}$
\be\label{e12}
\frac{3}{2}=\frac{1}{N}\sum_{\bk}
\frac{1}{1-|W_{\bk}/y_{0}|^{2}}.
\ee
Now that an appropriate self-avoiding loop can always be defined
out of a general one.  To see this, let us start a walk at a given site,
say, $j_{0}$.  Whenever a previous site is hit, we start again the walk
at that site and attribute the close loop to the renormalized $1/y$ of
the site. The procedure can be continued till the end of the original loop.
As a result, $1/y$ for a self-avoiding loop of length $(n+1)$  is
\be\label{e13}
\frac{1}{y}=\left\{\prod_{i=0}^{n}\left[\frac{1}{y_{0}}+\frac{1}{2}\times
\sum\mbox{loops excluding \{$j=0,\cdots,i-1$\}}
\right]\right\}^{1/(n+1)}
\cong \frac{3}{2y_{0}},
\ee
where (\ref{e12}) has been used and the last step comes from the
observation that the total number of sites are much larger than
those excluded.  Eq. (\ref{e13}) means that one can relax the
self-avoiding condition while replacing the correct $y$ by $y_{0}$.
Another way of seeing this is as
follows.  There are infinite number of self-closing loops starting
at site $j_{0}$ when the constraint is relaxed.  Let a single-loop
contribution be $L_{0}$,  we can define the renormalized one by putting the
rest
into $1/y$.  Since $L_{0}+L_{0}^{2}+\cdots=1$, $L_{0}=2/3$ and
the correct $L=1$, we obtain $1/y=3/(2y_{0})$.

The spin-spin correlations can be similarly considered.
We relax finding sites $i$ and $j$ on the same loop
to having two independent non-self-avoiding loops starting from
sites $i$ to $j$ and returning from $j$ to $i$. The over-counting
can be corrected again by the renormalization. But,
comparing the present case to the above one, a new over-counting
appears at site $j$, apart from those included in the renormalized $1/y$.
The final result should be multiplied by $2/3$ to correct it (the
overlap between the two paths are not considered in the above discussion).

This gives[\ref{r8}]
\be\label{e14}
<\hat{{\mathbf S}}_{A}(0)\cdot\hat{{\mathbf S}}_{B}({\mathbf R})>
=-\left|\frac{1}{N}\sum_{\bk}\frac{\exp(-i\bk\cdot{\mathbf R})
}{1-W_{\bk}/y_{0}}\right|^{2}.
\ee
where $W_{\bk}|_{\bk\ra\bk+\pi(1,1)}=-W_{\bk}$, which
assures no bonds between  same sublattices, has
be used. The correlations between same sublattices
turn out to have the same final
expression except for the sign factor (cf. [\ref{r7}]).

Let us briefly review the previous analysis[\ref{r8}]
that shows the long-range antiferromagnetic correlations.
Rewrite (\ref{e12}) in the form,
\[\frac{3}{2}=\frac{2}{\pi^{2}}\int_{-1}^{1}d\gamma
\frac{K(\sqrt{1-\gamma^{2}})}{1-W(\gamma)/y_{0}},\Sp
W(\gamma_{\bk})=W_{\bk},\]
where $K(x)$ is the complete elliptic integral of the first kind.

Consider now the limit $\eta\cong 1$, so that $W'(1)$ is very
large. The equality holds only if $y$ approaches $W(1)$ from
above so that the logarithmic divergence plays a role. In fact,
$y_{0}\cong W(1)+\Lambda\exp[-n_{0}W'(1)\pi/2]$,
with $\Lambda\sim o(1)$ being a cutoff and $n_{0}=0.607$.
For large ${\mathbf R}$, the spin-spin correlations take the form
$\sim (\xi/R)\exp(-R/\xi)$ with
\[\xi=(1/4)[W'(1)/\Lambda]^{1/2}\exp[n_{0}\pi W'(1)/4].\]
At $\eta=1$, $W'(1)=\infty$ leads to a condensation
 in (\ref{e14}) at $\bk=(0,0)$, thus long-range correlations.
Picking up this contribution gives
the staggered magnetization $M_{s}$
\[M_{s}=\left.\sqrt{-<\hat{{\mathbf S}}_{A}(0)\cdot\hat{{\mathbf
S}}_{B}({\mathbf R})>_{0}}
\right|_{R\rightarrow\infty}=
\frac{n_{0}}{2}\cong 0.303.\]
The ground-state energy is determined by the nearest neighbor
correlations. Including again the condensation, we find
\[E_{bond}=-J\left[1-\int_{0}^{1}\frac{2d\gamma}{\pi^{2}}\sqrt{1-\gamma^{2}}
 K(\sqrt{1-\gamma^{2}})\right]^{2}=-0.3352J.\]
Thus, the approximation recovers the conventional
spin-wave theory using the Holstein-Primakoff transformation
(expanding the square root to the order $1/S$).

The above argument for the ordering of the system can be extended
to general cases. Let max$(|W_{\bk}|)=|W_{\bk_{0}}|$.
It is clear that the system posses a long-range order if and only
if [$\bar{W}$ is equivalent to $W'(1)$ for isotropic $W_{\bk}$ near
$\bk_{0}$;
$\alpha,\delta=x,y$]
\[\bar{W}=\mbox{Det}\left(\frac{\partial^{2} |W_{\bk}|}{\partial
k_{\alpha}
\partial k_{\delta}}\right)_{\bk=\bk_{0}}\rightarrow \infty
\makebox[1in]{and}
\left.\frac{1}{N}\sum_{\bk}\frac{1}{1-|W_{\bk}/y_{0}|^{2}}\right|_{y_{0}\ra
|W_{\bk_{0}}|+0^{+}}<\frac{3}{2}.\]
For $W_{ij}$ falling  exponentially at large distances, second
derivatives of $|W_{\bk}|$ should be everywhere well-defined. The system is
thus
short-range correlated. In general, singular $W_{\bk}$ will leads to
non-exponential decays, such as power-law decays of $W_{ij}$.
It may end up with ordered
states.  For example, (\ref{e8}) decays as $R^{-2}$.
This result covers the general features found in numerical
studies[\ref{r9}].

\subsection{The proof of the optimization}

The RVB state of (\ref{e8}) is,  in fact,
  the optimized one of this class  within the analytic
self-consistency
approximation.  To show this, we incorporate (\ref{e12}) into
(\ref{e14}) with a lagrangian multiplier $\lambda_{L}$.  Let
$f_{\bk}=W_{\bk}/y_{0}$, we are led to maximize the expression
\[\sum_{\bk}\left\{(\gamma_{\bk}f_{\bk}-\lambda_{L})/(1
-f_{\bk}^{2})\right\}\],
where we have assumed, without loss of generality,
$f_{\bk}=f_{-\bk}=f_{\bk}^{\ast}$. The solutions of $f_{\bk}$ read,

\be\label{e15}
f_{\bk}=\frac{\lambda_{L}}{\gamma_{\bk}}\pm
\sqrt{(\frac{\lambda_{L}}{\gamma_{\bk}})^{2}-1}
\ee
Picking up the $|f_{\bk}|\leq 1$ solution leads to the one given
by (\ref{e8}) with $\eta=y_{0}/\lambda_{L}$.  It is straightforward
to show that $\eta=1$ minimizes the ground-state energy as expected.

The energy can also be rigorously evaluated without relaxing
the self-avoiding restriction. This is done via the Monte Carlo
method (cf. below for details). It yields, on a $48\times 48$
lattice, $-0.3344J$/bond and 0.311 for the
energy and the staggered magnetization (all digits are reliable).
This strongly justifies the analytic approximation.
There is no adjustable parameter.  The energy is indeed the
best presented in [\ref{r9}].  Comparing these values to the best
estimates $-0.3346(1)J$/bond and $0.31(2)$, cf. [\ref{r3}],
our approach may have virtually reproduced the exact ground state.

\subsection{The excited states}

Our ground state is rotationally  invariant.   As a result,
$<\Phi_{G}|\hat{{\mathbf{S}}}|\Phi_{G}>=0$.  Therefore the states
$\hat{S}_{j,\alpha}|\Phi_{G}>$, $\alpha=x,y,z$ are the excited
states of the system. More explicitly, $\hat{S}_{j,x}|\Phi_{G}>$ in
sublattice A can be written as,
\[ |\phi_{j,x}>=(\hat{b}^{\d}_{j\ua}\hat{b}_{j\da}
+\hat{b}^{\d}_{j\da}\hat{b}_{j\ua})|\Phi_{G}>=
\hat{B}_{j,x}^{\d}|\Phi_{G}>\]
where $\hat{B}_{j,x}^{\d}$ flips the ``spins'' of site $j$
so that different spin components are mixed.
It is easy to see that the three branches of states are orthogonal
to each others, but not among their own kinds.
We therefore construct the Bloch states
\be\label{e16}
|\phi_{\bk,\alpha}>=\frac{1}{N^{1/2}}\sum_{j}\exp(i\bk\cdot\br)
|\phi_{j,\alpha}>,\Sp \epsilon_{\bk,\alpha}=
\frac{<\phi_{\bk,\alpha}|\hat{H}|\phi_{\bk,\alpha}>}{
<\phi_{\bk,\alpha}|\phi_{\bk,\alpha}>}-E_{G}.
\ee
The calculation of the excitation spectrum reduces to the
evaluations of the
matrix elements $<\phi_{k,\alpha}|\phi_{l,\alpha}>$,
$<\phi_{k,\alpha}|\hat{{\mathbf S}}_{i}\cdot\hat{{\mathbf
S}}_{j}|\phi_{l,\alpha}>$.
By symmetry, $\epsilon_{\bk,x}=\epsilon_{\bk,y}=\epsilon_{\bk,z}=
\epsilon_{\bk}$ (but see below).
The computations can be most easily carried out via our matrix
method:  we simply need
to modify the transfer matrices at sites $i,j,k,l$.
The procedure for the $x$-branch is sketched below.

Consider first the four point terms in which $i,j,k,l$ are
different. The extra matrices at $i,j$ (the spin sites) have been given by
(\ref{e11}). The extra matrices at $k,l$ (the excitation sites) are
simply,
\be\label{e17}
\hat{B}_{k,x},\hat{B}^{\d}_{l,x}
\longrightarrow\left(\begin{array}{cc} {\mathbf 0} & {\mathbf I}\\
 {\mathbf I}& {\mathbf 0}\end{array}\right)_{k,l}.
\ee
There are cases where some of the sites coincide. We shall classify
them as three-point and two-point terms.  The corresponding matrices can
be easily deduced from (\ref{e11}) and (\ref{e17}) by multiplications of two
or more matrices at one site. One needs to pay some attentions to
the orders of the matrices:  The relevant operator to be averaged
is $\hat{B}_{k,x}\hat{{\mathbf S}}_{i}
\cdot\hat{{\mathbf S}}_{j}\hat{B}^{\d}_{l,x}$, with $\hat{\rho}=
|\Phi_{G}><\Phi_{G}|$ (which provides the ``loop gas'').
As an example, the three-point term with $k=i$ have
the matrices of $i,j$  which are simply (\ref{e11}) left-multiplied
by the matrix of $\hat{B}_{k,x}$ in (\ref{e17}),
\[   
-\frac{1}{4}\left(\begin{array}{cc} {\mathbf 0} & -{\mathbf I} \\  {\mathbf I}&
{\mathbf 0}
\end{array}\right)_{i}
\left(\begin{array}{cc} {\mathbf I} & {\mathbf 0}\\ {\mathbf 0}& -{\mathbf I}
\end{array}\right)_{j}
-\frac{1}{2}\left[
\left(\begin{array}{cc} {\mathbf I} & {\mathbf 0}\\ {\mathbf 0}& {\mathbf 0}
\end{array}\right)_{i}
\left(\begin{array}{cc} {\mathbf 0} & {\mathbf I}\\ {\mathbf 0}& {\mathbf 0}
\end{array}\right)_{j}+
\left(\begin{array}{cc} {\mathbf 0} & {\mathbf 0}\\ {\mathbf 0}& {\mathbf I}
\end{array}\right)_{i}
\left(\begin{array}{cc} {\mathbf 0} & {\mathbf 0}\\ {\mathbf I}& {\mathbf 0}
\end{array}\right)_{j}\right].\]
One can likewise consider the remaining cases.
Inserting these matrices at $i,j,k,l$,
one then computes the modified $\tilde{Y}_{N}$.
The non-zero loops involving the four sites are shown in Figure 1,
where the values attached are the ratios of their  contributions to
those without the inserting matrices.
They are used in the Monte Carlo evaluations.

In our calculation, the loop configurations are updated by randomly
choosing a pair of next nearest neighbor sites and exchanging their
loop connections with a probability satisfying the detailed balance
condition[\ref{r9}].  The ground-state energy per bond
is obtained by sampling over nearest neighbor bonds on same loops
(with a weight $-3/4$).   To find the excited energies, we randomly pick
up a spin bond $i,j$ and the first excited site $k$.  Summing the  sites
$l$ over the whole lattice for  a given $\bk$ can be managed  as
follows: Find the loop containing $k$.  Only the sites on this
loop contribute to the denominator in (\ref{e16}). This
algorithm ensures the positivity of the  denominator except for
$\bk=(\pi,\pi)$ where it is rigorously zero (thus the state is not
well defined). To obtain the numerator,
 classify the location of the bond into three cases: a) Both $i,j$
on the loop; b) Both $i,j$ not on the loop; c) One of $i,j$, say $i$,
 on the loop.  In cases a) and b), again
only the sites on the loop of $k$ contribute, while
in case c) the sites on the loop containing $j$ are considered.
In all the cases, the weights in Figure 1 are used as mentioned.
Figure 2 presents the numerical result (plotted in $\diamond$)
on a $10\times 10$ lattice. Two points are immediately clear: (1) The
gap at $\bk=(0,0)$ is virtually zero.  (2) The spectrum is linear
at small $\bk$, in agreement with some of the well-established
results. The technical difficulty here is that one needs at least the
precise value of the finite-size ground-state energy  within a relative
error of $10^{-4}$.  Note that, in contrast, the dimer state
with only the nearest neighbor RVB bond
always gives a negative energy gap at $\bk=0$ (beyond the error
range). Thus the latter is, as a matter of fact, not a suitable trial
ground state.

The above spectrum does not agree {\em in details} with those
obtained via more sophisticated spin-wave or related theories[\ref{r3}].
We now discuss the breaking of the rotational symmetry
which is, we suspect, the source of the discrepancies.
 At zero temperature, it is generally believed that the system
should suffer a symmetry breaking such that the spins like to align
up on one sublattice and down on the other. When this happens, different
branches of the excited states proposed above are no longer
orthogonal. One is therefore  forced to reconstruct  the excited states.
The details might be quite complicated and we shall explore it
elsewhere. How is this going to show up in the above calculation?
A natural guess is that, since the periodicity of the system is
reduced, we only sample over terms with $k,l$ belonging to same sublattices.

We stress that this is based on our intuitive understanding of the
system and has not been proved.   The result (plotted in $\bullet$)
is also shown in Figure 2.
It turns out to fit the renormalized spin-wave result
$\epsilon_{\bk}=2JZ_{r}\sqrt{1-\gamma_{\bk}^{2}}$, with $Z_{r}\cong
1.23$, in excellent agreement with  other results (cf., in particular,
[\ref{r12}] done on a supercomputer!).
This greatly narrows the differences between the RVB and the
spin-wave approaches, although details remain to be clarified.

In conclusion, we have obtained in this work
an optimized, parameter-free  RVB state for the 2-d
antiferromagnetic Heisenberg model.
We have also developed effective methods for
calculations of various physical quantities, in particular, the
excitation spectrum of the ground state (which has been a long-standing
difficulty in this context).  Our results agree with, and in some aspects, are
better than the conventional spin-wave theory.

This work was supported by the National Science Council
and the National Educational Committee of China.

\renewcommand{\thesubsection}{}
\subsection{References}

\begin{enumerate}
\item     For representative discussions of high-$T_{c}$ theories
	and experiments,
       see P. W. Anderson and R. Schrieffer,  Physics Today \u{44}, 55
	     (1991); B. Batlogg, {\em ibid} \u{44}, 44 (1991).
     \label{r1}
\item     For a recent comprehensive summary on the Heisenberg model, see
     E. Manousakis, Rev. Mod. Phys. \u{63}, 1 (1991). \label{r3}
\item     D. P. Arovas and A. Auerbach, Phys. Rev. B \u{38}, 316 (1988);
     A. Auerbach and D. P. Arovas, J. Appl. Phys. \u{67}, 5734 (1990).
     \label{r5}
\item	  S. Sarker, C. Jayaprakash, H. R. Krishamurthy, and M. Ma,
	Phys. Rev. B \u{40}, 5028 (1989). \label{r5a}
\item 	  C. Jayaprakash, H. R. Krishnamurthy, and S. Sarker, Phys. Rev. B
	\u{40}, 2610 (1989); C. L. Kane, P. A. Lee, and T. K. Ng,
	B. Chakraborty, and N. Read, Phys. Rev. B \u{41}, 2653 (1990).
	\label{r5b}
\item     Y.-C. Chen, Physica C \u{202}, 345 (1992).
     \label{r6}
\item     Y.-C. Chen, Physica C \u{204}, 88 (1992). \label{r7}
\item     Y.-C. Chen, Phys. Lett. A \u{174}, 329 (1993).\label{r8}
\item     S. Liang, B. Doucot, and P. W. Anderson, Phys. Rev. Lett.
     \u{61}, 365 (1988).
     \label{r9}
\item     B. Sutherland, Phys. Rev. B \u{37}, 3786 (1988); {\em ibid}
     \u{38}, 6855 (1988). \label{r10}
\item     G. Chen, H.-Q. Ding, and W. A. Goddard III, Phys. Rev. B \u{46},
     2933 (1992). 
     \label{r12}
\end{enumerate}
\newpage
\subsection{Figure captions}
\begin{description}
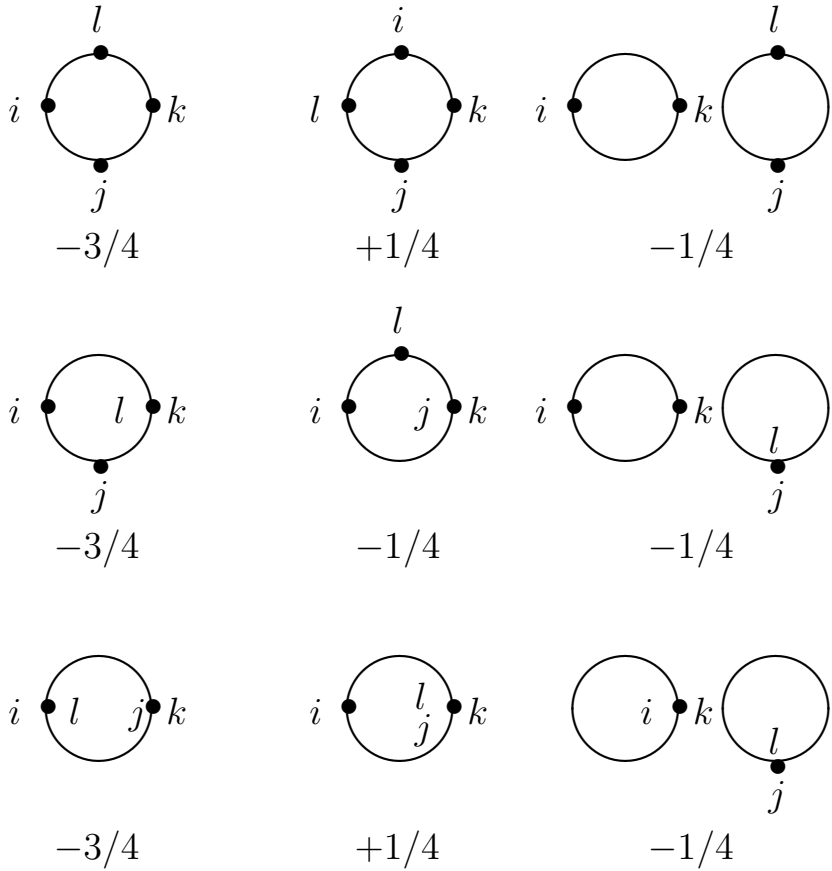
\item[Fig. 1] A list of the non-zero loops involving the four sites
$i,j,k,l$
     and their contributions.
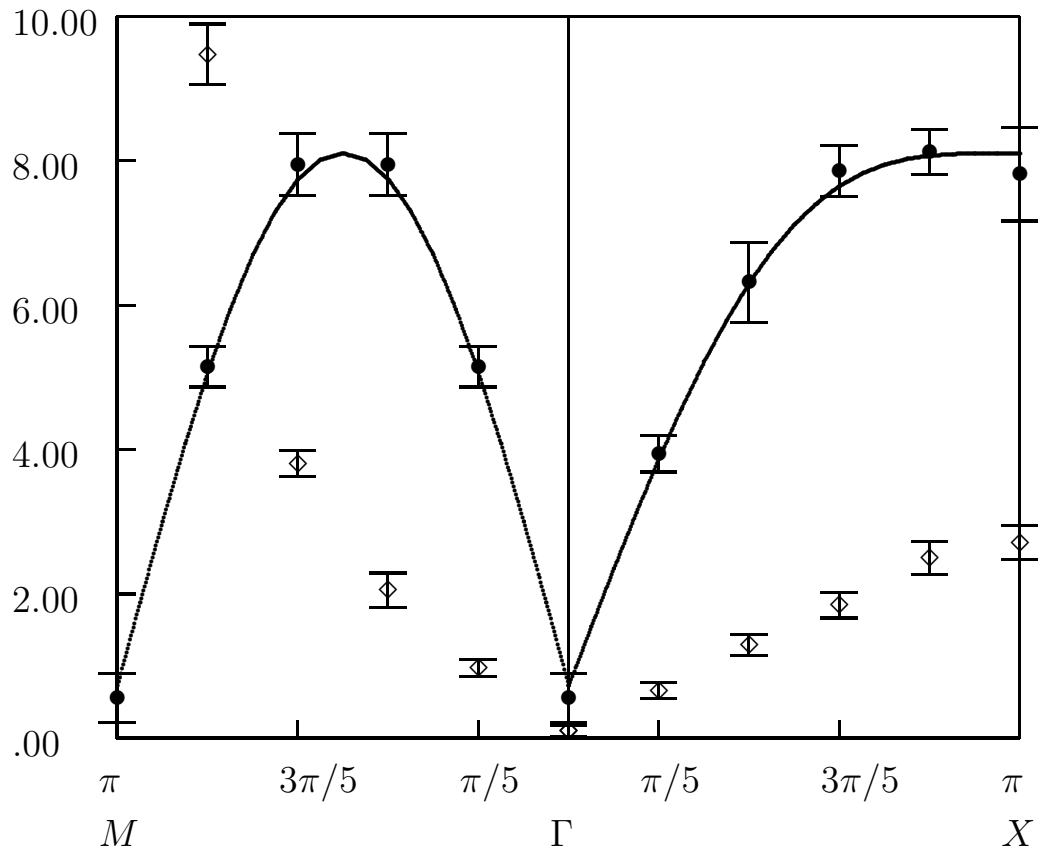
\item[Fig. 2]  The excitation spectrum $\epsilon_{\bk}/J$, plotted in
     $\diamond$, along the (1,0) axis (on the right-hand side) and the (1,1)
     direction.  The gap at $\bk=(0,0)$ is virtually zero.
     Plotted in $\bullet$ is $\epsilon_{\bk}/J$ ($\times 3$)
     upon breaking of the rotational symmetry. Solid curve ($\times 3$),
     with an upward shift $\cong 0.24$ (suppressed
     as $L$ increases),  is the renormalized spin-wave result
     with $Z_{r}=1.23$.
\end{description}

\newpage
  \large
  \renewcommand{\thepage}{Figure 1}
  \setlength{\unitlength}{1mm}
  \begin{picture}(100,100)(-50,-15)
  \thicklines


\put(-40,40){\circle{20}}
\put(-34,39){$\bullet$}\put(-31,38){$k$}
\put(-41,46){$\bullet$}\put(-41,50){$l$}
\put(-48,39){$\bullet$}\put(-52,38){$i$}
\put(-41,31){$\bullet$}\put(-41,27){$j$}
\put(-46,20){$-3/4$}

\put(0,40){\circle{20}}
\put(6,39){$\bullet$}\put(9,38){$k$}
\put(-1,46){$\bullet$}\put(-1,50){$i$}
\put(-8,39){$\bullet$}\put(-12,38){$l$}
\put(-1,31){$\bullet$}\put(-1,27){$j$}
\put(-6,20){$+1/4$}

\put(30,40){\circle{20}}
\put(36,39){$\bullet$}\put(39,38){$k$}
\put(22,39){$\bullet$}\put(18,38){$i$}
\put(50,40){\circle{20}}
\put(49,46){$\bullet$}\put(49,50){$l$}
\put(49,31){$\bullet$}\put(49,27){$j$}
\put(33,20){$-1/4$}

\put(-40,0){\circle{20}}
\put(-34,-1){$\bullet$}\put(-31,-2){$k$}
\put(-38,-2){$l$}
\put(-48,-1){$\bullet$}\put(-52,-2){$i$}
\put(-41,-9){$\bullet$}\put(-41,-13){$j$}
\put(-46,-20){$-3/4$}

\put(0,0){\circle{20}}
\put(6,-1){$\bullet$}\put(9,-2){$k$}
\put(-1,6){$\bullet$}\put(-1,10){$l$}
\put(-8,-1){$\bullet$}\put(-12,-2){$i$}
\put(2,-2){$j$}
\put(-6,-20){$-1/4$}

\put(30,0){\circle{20}}
\put(36,-1){$\bullet$}\put(39,-2){$k$}
\put(22,-1){$\bullet$}\put(18,-2){$i$}
\put(50,0){\circle{20}}
\put(49,-6){$l$}
\put(49,-9){$\bullet$}\put(49,-13){$j$}
\put(33,-20){$-1/4$}


\put(-40,-40){\circle{20}}
\put(-34,-41){$\bullet$}\put(-31,-42){$k$}
\put(-44,-42){$l$}
\put(-48,-41){$\bullet$}\put(-52,-42){$i$}
\put(-36,-42){$j$}
\put(-46,-60){$-3/4$}

\put(0,-40){\circle{20}}
\put(6,-41){$\bullet$}\put(9,-42){$k$}
\put(2,-40){$l$}
\put(-8,-41){$\bullet$}\put(-12,-42){$i$}
\put(2,-44){$j$}
\put(-6,-60){$+1/4$}

\put(30,-40){\circle{20}}
\put(36,-41){$\bullet$}\put(39,-42){$k$}
\put(32,-42){$i$}
\put(50,-40){\circle{20}}
\put(49,-46){$l$}
\put(49,-49){$\bullet$}\put(49,-53){$j$}
\put(33,-60){$-1/4$}

\end{picture}

	\newpage
  \renewcommand{\thepage}{Figure 2}
  \setlength{\unitlength}{1.2mm}
  \begin{picture}(120,100)(-10,-0)
  \thicklines
  \put(51,0){\line(0,1){80}}
  \put(1,80){\line(1,0){100}}
  \put(1,0){\line(1,0){100}}
  \put(1,0){\line(0,1){80}}
  \put(101,0){\line(0,1){80}}
  \put(49,-12){$\Gamma$}
  \put(-1,-12){$M$}
  \put(99,-12){$X$}
 \put(    -1,    -6){$\pi$}
 \put(    19,    -6){$3\pi/5$}
 \put(    21,     0){\line(0,1){2}}
 \put(    39,    -6){$\pi/5$}
 \put(    41,     0){\line(0,1){2}}
 \put(    59,    -6){$\pi/5$}
 \put(    61,     0){\line(0,1){2}}
 \put(    79,    -6){$3\pi/5$}
 \put(    81,     0){\line(0,1){2}}
 \put(    99,    -6){$\pi$}
 \put(   -12,    -2){   .00}
 \put(   -12,    14){  2.00}
 \put(     1,    16){\line(1,0){2}}
 \put(   -12,    30){  4.00}
 \put(     1,    32){\line(1,0){2}}
 \put(   -12,    46){  6.00}
 \put(     1,    48){\line(1,0){2}}
 \put(   -12,    62){  8.00}
 \put(     1,    64){\line(1,0){2}}
 \put(   -12,    78){ 10.00}
 \put(  50.00,   -.21){$\diamond$}
 \put(  51.00,    .19){\line(0,1){   1.19}}
 \put(  49.00,    .19){\line(1,0){4}}
 \put(  49.00,   1.38){\line(1,0){4}}
 \put(  60.00,   4.26){$\diamond$}
 \put(  61.00,   4.39){\line(0,1){   1.75}}
 \put(  59.00,   4.39){\line(1,0){4}}
 \put(  59.00,   6.14){\line(1,0){4}}
 \put(  70.00,   9.34){$\diamond$}
 \put(  71.00,   9.19){\line(0,1){   2.31}}
 \put(  69.00,   9.19){\line(1,0){4}}
 \put(  69.00,  11.50){\line(1,0){4}}
 \put(  80.00,  13.72){$\diamond$}
 \put(  81.00,  13.30){\line(0,1){   2.84}}
 \put(  79.00,  13.30){\line(1,0){4}}
 \put(  79.00,  16.14){\line(1,0){4}}
 \put(  90.00,  18.95){$\diamond$}
 \put(  91.00,  18.09){\line(0,1){   3.73}}
 \put(  89.00,  18.09){\line(1,0){4}}
 \put(  89.00,  21.82){\line(1,0){4}}
 \put( 100.00,  20.66){$\diamond$}
 \put( 101.00,  19.77){\line(0,1){   3.77}}
 \put(  99.00,  19.77){\line(1,0){4}}
 \put(  99.00,  23.54){\line(1,0){4}}
 \put(  50.00,   -.21){$\diamond$}
 \put(  51.00,    .19){\line(0,1){   1.19}}
 \put(  49.00,    .19){\line(1,0){4}}
 \put(  49.00,   1.38){\line(1,0){4}}
 \put(  40.00,   6.75){$\diamond$}
 \put(  41.00,   6.83){\line(0,1){   1.84}}
 \put(  39.00,   6.83){\line(1,0){4}}
 \put(  39.00,   8.67){\line(1,0){4}}
 \put(  30.00,  15.41){$\diamond$}
 \put(  31.00,  14.51){\line(0,1){   3.78}}
 \put(  29.00,  14.51){\line(1,0){4}}
 \put(  29.00,  18.30){\line(1,0){4}}
 \put(  20.00,  29.45){$\diamond$}
 \put(  21.00,  28.99){\line(0,1){   2.91}}
 \put(  19.00,  28.99){\line(1,0){4}}
 \put(  19.00,  31.90){\line(1,0){4}}
 \put(  10.00,  74.80){$\diamond$}
 \put(  11.00,  72.44){\line(0,1){   6.72}}
 \put(   9.00,  72.44){\line(1,0){4}}
 \put(   9.00,  79.16){\line(1,0){4}}

\large
 \put(  50.50,   4.76){$\cdot$}
 \put(  53.00,  11.30){$\cdot$}
 \multiput( 50.50,  4.76)(    .1250,    .3270){  20}{$\cdot$}
 \put(  55.50,  17.74){$\cdot$}
 \multiput( 53.00, 11.30)(    .1250,    .3220){  20}{$\cdot$}
 \put(  58.00,  23.98){$\cdot$}
 \multiput( 55.50, 17.74)(    .1250,    .3121){  20}{$\cdot$}
 \put(  60.50,  29.94){$\cdot$}
 \multiput( 58.00, 23.98)(    .1250,    .2977){  20}{$\cdot$}
 \put(  63.00,  35.52){$\cdot$}
 \multiput( 60.50, 29.94)(    .1250,    .2791){  20}{$\cdot$}
 \put(  65.50,  40.66){$\cdot$}
 \multiput( 63.00, 35.52)(    .1250,    .2570){  20}{$\cdot$}
 \put(  68.00,  45.30){$\cdot$}
 \multiput( 65.50, 40.66)(    .1250,    .2320){  20}{$\cdot$}
 \put(  70.50,  49.40){$\cdot$}
 \multiput( 68.00, 45.30)(    .1250,    .2049){  20}{$\cdot$}
 \put(  73.00,  52.93){$\cdot$}
 \multiput( 70.50, 49.40)(    .1250,    .1766){  20}{$\cdot$}
 \put(  75.50,  55.89){$\cdot$}
 \multiput( 73.00, 52.93)(    .1250,    .1480){  20}{$\cdot$}
 \put(  78.00,  58.29){$\cdot$}
 \multiput( 75.50, 55.89)(    .1250,    .1201){  20}{$\cdot$}
 \put(  80.50,  60.16){$\cdot$}
 \multiput( 78.00, 58.29)(    .1250,    .0937){  20}{$\cdot$}
 \put(  83.00,  61.56){$\cdot$}
 \multiput( 80.50, 60.16)(    .1250,    .0696){  20}{$\cdot$}
 \put(  85.50,  62.53){$\cdot$}
 \multiput( 83.00, 61.56)(    .1250,    .0488){  20}{$\cdot$}
 \put(  88.00,  63.16){$\cdot$}
 \multiput( 85.50, 62.53)(    .1250,    .0316){  20}{$\cdot$}
 \put(  90.50,  63.53){$\cdot$}
 \multiput( 88.00, 63.16)(    .1250,    .0183){  20}{$\cdot$}
 \put(  93.00,  63.71){$\cdot$}
 \multiput( 90.50, 63.53)(    .1250,    .0091){  20}{$\cdot$}
 \put(  95.50,  63.78){$\cdot$}
 \multiput( 93.00, 63.71)(    .1250,    .0035){  20}{$\cdot$}
 \put(  98.00,  63.80){$\cdot$}
 \multiput( 95.50, 63.78)(    .1250,    .0008){  20}{$\cdot$}
 \put( 100.50,  63.80){$\cdot$}
 \multiput( 98.00, 63.80)(    .1250,    .0001){  20}{$\cdot$}
 \put(  50.50,   4.76){$\cdot$}
 \put(  48.00,  14.00){$\cdot$}
 \multiput( 50.50,  4.76)(   -.1250,    .4618){  20}{$\cdot$}
 \put(  45.50,  23.00){$\cdot$}
 \multiput( 48.00, 14.00)(   -.1250,    .4504){  20}{$\cdot$}
 \put(  43.00,  31.56){$\cdot$}
 \multiput( 45.50, 23.00)(   -.1250,    .4280){  20}{$\cdot$}
 \put(  40.50,  39.46){$\cdot$}
 \multiput( 43.00, 31.56)(   -.1250,    .3950){  20}{$\cdot$}
 \put(  38.00,  46.51){$\cdot$}
 \multiput( 40.50, 39.46)(   -.1250,    .3522){  20}{$\cdot$}
 \put(  35.50,  52.52){$\cdot$}
 \multiput( 38.00, 46.51)(   -.1250,    .3008){  20}{$\cdot$}
 \put(  33.00,  57.37){$\cdot$}
 \multiput( 35.50, 52.52)(   -.1250,    .2420){  20}{$\cdot$}
 \put(  30.50,  60.91){$\cdot$}
 \multiput( 33.00, 57.37)(   -.1250,    .1773){  20}{$\cdot$}
 \put(  28.00,  63.07){$\cdot$}
 \multiput( 30.50, 60.91)(   -.1250,    .1081){  20}{$\cdot$}
 \put(  25.50,  63.80){$\cdot$}
 \multiput( 28.00, 63.07)(   -.1250,    .0363){  20}{$\cdot$}
 \put(  23.00,  63.07){$\cdot$}
 \multiput( 25.50, 63.80)(   -.1250,   -.0363){  20}{$\cdot$}
 \put(  20.50,  60.91){$\cdot$}
 \multiput( 23.00, 63.07)(   -.1250,   -.1081){  20}{$\cdot$}
 \put(  18.00,  57.37){$\cdot$}
 \multiput( 20.50, 60.91)(   -.1250,   -.1773){  20}{$\cdot$}
 \put(  15.50,  52.52){$\cdot$}
 \multiput( 18.00, 57.37)(   -.1250,   -.2420){  20}{$\cdot$}
 \put(  13.00,  46.51){$\cdot$}
 \multiput( 15.50, 52.52)(   -.1250,   -.3008){  20}{$\cdot$}
 \put(  10.50,  39.46){$\cdot$}
 \multiput( 13.00, 46.51)(   -.1250,   -.3522){  20}{$\cdot$}
 \put(   8.00,  31.56){$\cdot$}
 \multiput( 10.50, 39.46)(   -.1250,   -.3950){  20}{$\cdot$}
 \put(   5.50,  23.00){$\cdot$}
 \multiput(  8.00, 31.56)(   -.1250,   -.4280){  20}{$\cdot$}
 \put(   3.00,  14.00){$\cdot$}
 \multiput(  5.50, 23.00)(   -.1250,   -.4504){  20}{$\cdot$}
 \put(    .50,   4.76){$\cdot$}
 \multiput(  3.00, 14.00)(   -.1250,   -.4618){  20}{$\cdot$}

 \put(  50.00,   3.45){$\bullet$}
 \put(  51.00,   1.72){\line(0,1){   5.46}}
 \put(  49.00,   1.72){\line(1,0){4}}
 \put(  49.00,   7.18){\line(1,0){4}}
 \put(  60.00,  30.52){$\bullet$}
 \put(  61.00,  29.51){\line(0,1){   4.02}}
 \put(  59.00,  29.51){\line(1,0){4}}
 \put(  59.00,  33.53){\line(1,0){4}}
 \put(  70.00,  49.55){$\bullet$}
 \put(  71.00,  46.12){\line(0,1){   8.86}}
 \put(  69.00,  46.12){\line(1,0){4}}
 \put(  69.00,  54.98){\line(1,0){4}}
 \put(  80.00,  61.88){$\bullet$}
 \put(  81.00,  60.07){\line(0,1){   5.62}}
 \put(  79.00,  60.07){\line(1,0){4}}
 \put(  79.00,  65.69){\line(1,0){4}}
 \put(  90.00,  63.98){$\bullet$}
 \put(  91.00,  62.50){\line(0,1){   4.95}}
 \put(  89.00,  62.50){\line(1,0){4}}
 \put(  89.00,  67.45){\line(1,0){4}}
 \put( 100.00,  61.55){$\bullet$}
 \put( 101.00,  57.34){\line(0,1){  10.42}}
 \put(  99.00,  57.34){\line(1,0){4}}
 \put(  99.00,  67.75){\line(1,0){4}}
 \put(  40.00,  40.17){$\bullet$}
 \put(  41.00,  38.90){\line(0,1){   4.54}}
 \put(  39.00,  38.90){\line(1,0){4}}
 \put(  39.00,  43.44){\line(1,0){4}}
 \put(  30.00,  62.59){$\bullet$}
 \put(  31.00,  60.18){\line(0,1){   6.83}}
 \put(  29.00,  60.18){\line(1,0){4}}
 \put(  29.00,  67.00){\line(1,0){4}}
 \put(  20.00,  62.59){$\bullet$}
 \put(  21.00,  60.18){\line(0,1){   6.83}}
 \put(  19.00,  60.18){\line(1,0){4}}
 \put(  19.00,  67.00){\line(1,0){4}}
 \put(  10.00,  40.17){$\bullet$}
 \put(  11.00,  38.90){\line(0,1){   4.54}}
 \put(   9.00,  38.90){\line(1,0){4}}
 \put(   9.00,  43.44){\line(1,0){4}}
 \put(    .00,   3.45){$\bullet$}
 \put(   1.00,   1.72){\line(0,1){   5.46}}
 \put(  -1.00,   1.72){\line(1,0){4}}
 \put(  -1.00,   7.18){\line(1,0){4}}
  \end{picture}
  \end{document}